\documentclass[aps,preprint,showpacs,superscriptaddress,groupedaddress]{revtex4}  
\usepackage{graphicx}  
\usepackage{dcolumn}   
\usepackage{bm}        
\usepackage{amssymb}   

\usepackage{hyperref}
\usepackage{epstopdf}
\usepackage{subfigure}
\usepackage{pdfpages}

\usepackage{placeins} 
\usepackage{placeins} 
\usepackage{CJK}
\usepackage{indentfirst}

\usepackage{savesym}
\usepackage{amsmath}
\savesymbol{iint}
\usepackage{txfonts}
\restoresymbol{TXF}{iint}

\begin{document}

\title{Dependency-based targeted attacks in interdependent networks}

\author{Dong Zhou}
\affiliation{School of Reliability and Systems Engineering, Beihang University, Beijing 100191, China}
\affiliation{National Key Laboratory of Science and Technology on Reliability and Environmental Engineering, Beijing 100191, China}
\author{Amir Bashan}
\affiliation{Department of Physics, Bar-Ilan University, Ramat Gan 52900, Israel}


%

\maketitle

\section*{Abstract}

Modern large network systems normally work in cooperation and incorporate dependencies between their components for purposes of efficiency and regulation. Such dependencies may become a major risk since they can cause small scale failures to propagate throughout the system. Thus, the dependent nodes could be a natural target for malicious attacks that aim to exploit these vulnerabilities. Here, we consider for the first time a new type of targeted attacks that are based on the dependency between the networks. We study strategies of attacks that range from dependency-first to dependency-last, where a fraction $1-p$ of the nodes with dependency links, or nodes without dependency links, respectively, are initially attacked. We systematically analyze, both analytically and numerically, the percolation transition of partially interdependent Erd\H{o}s-R\'{e}nyi (ER) networks, where a fraction $q$ of the nodes in each network are dependent upon nodes the other network. We find that for a broad range of dependency strength $q$, `dependency-first' strategy, which intuitively is expected to increase the system's vulnerability, actually leads to a more stable system, in terms of lower critical percolation threshold $p_c$, compared with random attacks of the same size. In contrast, the `dependency-last' strategy leads to a more vulnerable system, i.e., higher $p_c$, compared with a random attack. By exploring the dynamics of the cascading failures initiated by dependency-based attacks, we explain this counter-intuitive effect. Our results demonstrate that the most vulnerable components in a system of interdependent networks are not necessarily the ones that lead to the maximal immediate impact but those which initiate a cascade of failures with maximal accumulated damage.

\section{Introduction}
The robustness of interdependent networks and multiplex networks has been studied extensively over the past decade~\cite{buldyrev2010catastrophic,parshani2010interdependent,buldyrev2011interdependent,Hu2011percolation,BaxterPRL2012,hu2013percolation,gao2011robustness,gao2012networks,li2012cascading,JianxiPRE2013,cellai2013percolation,bashan2013extreme,GohPRECorr2014,havlin2014vulnerability,BianconiPRE2014,ZhouDongPRE2014,YanqingNP2014,havlin2015percolation,Cellai2016PRE,Hackett2016PRX,Yuan2017PNAS,Radicchi2017}. This interest stems from the realization that many real-world network systems, especially critical infrastructure systems, are not isolated but rather interconnected and interdependent upon other networks. While in normal conditions the coupling is designed to improve the efficiency and regulation of the entire system, it also can become a risk factor when the system experiences failures. A first analytical framework to analyze the effect of interdependence between networks was provided by Buldyrev \emph{et al.}~\cite{buldyrev2010catastrophic} in 2010. They studied a model of two random networks where the functionality of nodes in each network is interdependent upon the functionality of nodes in the other network. In such systems, any failed node may trigger a cascade of failures in both networks. The resilience of the system is then represented by the size of the functional part of the network, $P_\infty$, after a failure of a fraction $1-p$ of the nodes (so $p$ represents the remaining part of the nodes after the initial failures), yielding a percolation transition. It was shown that the transition from a functional system, i.e., $P_\infty>0$, to complete collapse of the system, i.e., $P_\infty=0$, is discontinuous at certain $p=p_c$ \cite{buldyrev2010catastrophic}. Such abrupt transition represents a major vulnerability of the system.

A generalization of the analytical model to a case of partial dependence was proposed by Parshani \emph{et al.}~\cite{parshani2010interdependent}. It was shown that the coupling strength between the networks, represented by the fraction $q$ of interdependent nodes, affects two aspects of the way the system collapses~\cite{parshani2010interdependent,buldyrev2011interdependent,Hu2011percolation,hu2013percolation,gao2011robustness,gao2012networks,JianxiPRE2013}. First, it determines the critical threshold $p_c$ below which the system completely collapses. 
Second, the coupling strength determines the nature of the transition. For strong coupling, i.e., a high fraction of interdependent nodes, initial failures can initiate cascading failures that yield an abrupt collapse of the system, in the form of a first-order phase transition. Reducing the coupling strength below a critical value, $q_c$, leads to a change from an abrupt collapse to a continuous decrease of the size of the network, in the form of a second-order phase transition.

Real-world critical infrastructures are not only subject to random failures but may also be targets of malicious attacks that aim to maximize the damage to the entire system. In order to understand the vulnerabilities that arise from the coupling between network systems, models of targeted attacks on interdependent networks have been analyzed~\cite{Huang2011,Dong2012,Dong2013}. For example, Huang \emph{et al.} investigate the degree-based targeted attacks on fully interdependent Erd\H{o}s-R\'{e}nyi (ER) networks and scale-free (SF) networks, and they found that degree-based targeted attack problems can be mapped to random attack problems~\cite{Huang2011}. Dong \emph{et al.} further studied degree-based targeted attack in partially interdependent networks and show that there is a critical coupling strength, and when the coupling strength is larger than the critical value, the system shows a first-order transition~\cite{Dong2012}. In another related paper, their analytical results have been generalized to several types of networks of networks~\cite{Dong2013}. In 2017, Kleineberg \emph{et al.} showed that many realistic multiplex networks are surprisingly robust against degree-based targeted attacks, and geometric correlations have been found to be the major reason~\cite{Kleineberg2017PRL}.

Yet, the research of targeted attacks on interdependent networks focused on only one aspect of the system's vulnerability, namely, the role of the nodes within their individual network. However, the special vulnerability of interdependent networks primarily stems from the existence of dependencies between the networks. Particularly, the cascading failures are resulted by the synergy between processes within the individual networks, represented by connectivity links between nodes of the same network, and the interrelations between the networks, represented by dependency links between the networks. Therefore, malicious attacks could try to exploit the vulnerability arises from the dependency links. Indeed, attacking nodes with inter-dependencies seems to have an obvious advantage (from the point of view of the attacker): in addition to the nodes attacked directly, more failures will be caused due to their direct dependency relations. So far, the effect of such targeted attacks on interdependent networks has not been investigated.

To address this critical question, we set out to systematically study the dependency-based target attack strategies on partially interdependent ER networks. These dependency-based attacks can focus on nodes connected with dependency links, namely `dependency-first' attacks, or alternatively avoid those nodes and focus on nodes without dependency links, that is 'dependency-last' attacks. Using numerical simulations we analyzed the percolation transition resulted from each strategy for different values of the coupling strength $q$. We find that for each strategy there is a critical coupling strength $q^*$ such that for $q<q^*$ the resulted percolation transitions have a counter-intuitive behavior, where the system is more robust to 'dependency-first' attack strategies, i.e., smaller value of the percolation threshold $p_c$, compared with random attacks of the same size. In contrast, 'dependency-last' strategy results in a percolation transition with a higher value of $p_c$ compared with a random attack. In this manuscript, we study and explain this surprising behavior.



\section{Model description}
To study the resilience of interdependent networks, we study percolation transitions caused by targeted attacks on nodes with dependency links. To do this, we consider two partially interdependent networks $A$ and $B$, where a fraction of nodes, $q$, in $A$ and $B$ have dependency links. We assume that all the dependency links are bidirectional, and they fulfill the no-feedback condition: a node from $A$ or $B$ depends on no more than one node from the other network, and if a node $A_i$ depends on node $B_j$, and $B_j$ depends on $A_k$, then $i=k$~\cite{gao2012networks}.  The nodes with and without dependency links are called ``dependent nodes'' and ``non-dependent nodes'', respectively. An attacked node is represented by removing it from the network.

\begin{figure}
 \centering
 \subfigure{
  \includegraphics[width=0.65\textwidth,trim=10 10 10 10,clip=false]{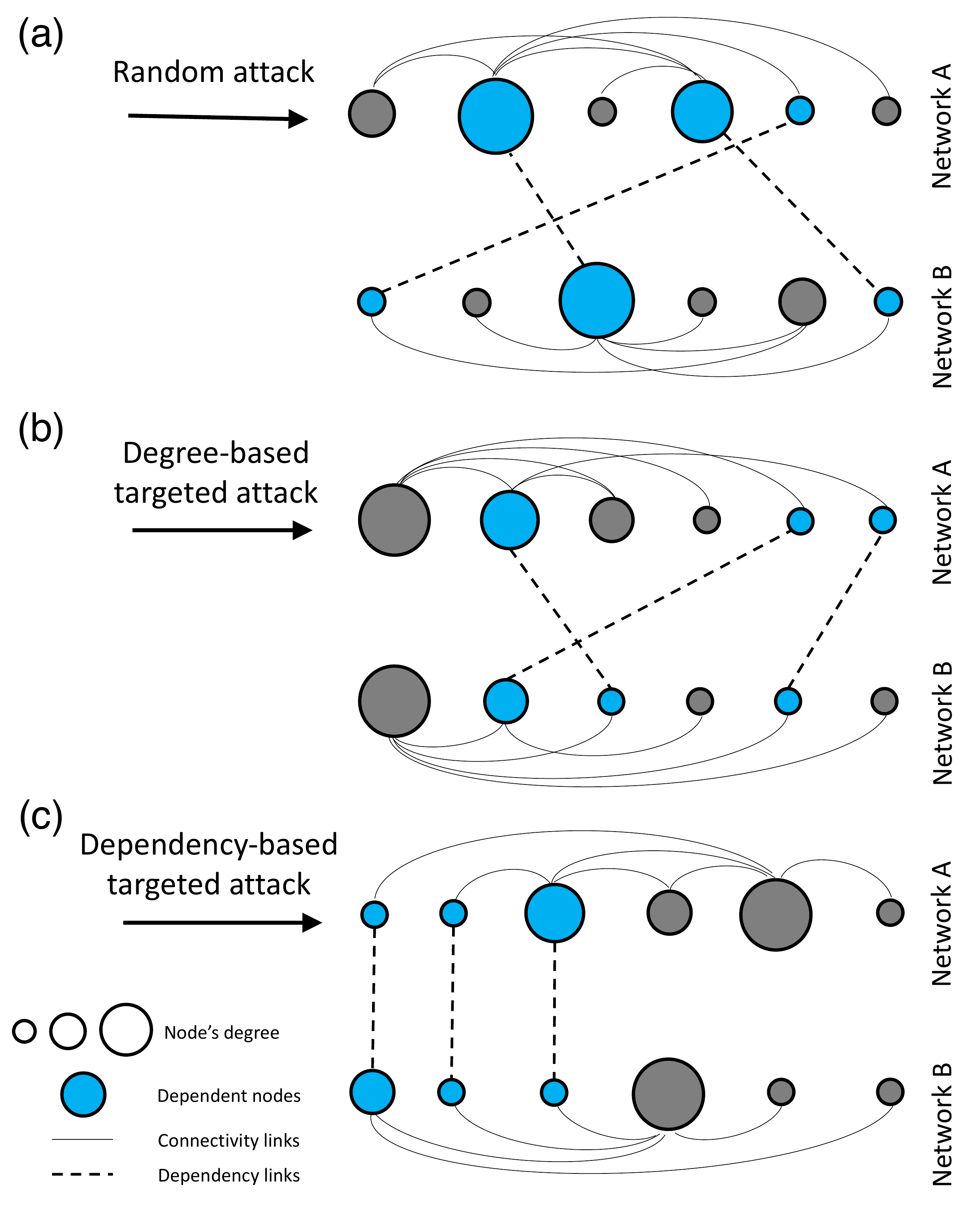}}
\caption{A schematic plot of partially interdependent networks, under three types of attacks: a) {\it Random attack}: Nodes are attacked at random, b) {\it Degree-based targeted attack}: The order of attacked nodes is determined by the number of connected nodes in the individual networks. c) {\it Dependency-based targeted attacks}: In "dependency-first" attack, nodes are attacked first if their failure will cause node in the other network to also fail, due to dependency relations. In "dependency-last" attack, these nodes are attacked lastly.}
 \label{fig0}
\end{figure}

We consider strategies for initial node attack that range between two opposite extremes:

{\bf \noindent `Dependency-first' targeted attack:} a fraction $1-p$ of nodes from $A$ are selected and removed, where dependent nodes are randomly selected first. If $(1-p)<q$, only dependent nodes will be removed. If $(1-p)>q$, all the dependency nodes will be removed and a fraction of $1-p-q$ of non-dependent nodes will be randomly removed.

{\bf \noindent `Dependency-last' targeted attack:} a fraction $1-p$ of nodes from $A$ are selected and removed, where non-dependent nodes are randomly selected first. If $p>q$, only non-dependent nodes will be removed. If $p<q$, all the non-dependency nodes will be removed and a fraction of $1-p-(1-q)$ of dependent nodes will be randomly removed.

An intermediate case between these two extremes is the standard 'random-attack'. In this case, nodes are randomly removed from both dependent and non-dependent nodes.

We define a general attack strategy with the parameter $\beta$ that includes as specific cases these two extreme attack strategies, as well as the case of a random attack. In this definition, a fraction of $1-p$ of nodes are initially removed, but the probability of choosing a dependent node is $\alpha$ times the probability of choosing a non-dependent node, where $\alpha \equiv (1+\beta)/(1-\beta)$. Here, $\beta\in[-1,1]$ is used as a parameter that modulates the ratio of  dependent nodes in the initial attack. The cases of $\beta=1$ and $\beta=-1$ correspond to dependency-first and dependency-last targeted attacks, respectively, and $\beta=0$ corresponds to a random attack. Attack strategies with a higher probability to attack dependent nodes, i.e., $0<\beta<1$, are referred to as `Dependency-biased' attacks, while attacks with a higher probability to attack non-dependent nodes, i.e., $-1<\beta<0$, are referred to as `Non-dependency-biased' attacks.

We compare our results to the traditional `degree-based' attacks, namely, `high-degree first' and `low-degree first' for which a fraction $1-p$ of nodes from $A$ are selected and removed, where nodes are sorted according to their degrees, and chosen one by one from the largest- or smallest-degree nodes, respectively~\cite{Huang2011,Dong2012,Dong2013}. Obviously, given the very different nature of `degree-based' and `dependency-based' attacks, their impact on the networks is quantitatively different. For example, attacking a single `hub', instead of just a random node, can dramatically increase the immediate damage and affect many other nodes in its network, while attacking a single `dependent-node' causes additional removal of only one dependent-node in the other network. Nevertheless, by systematic analyzing each of these attack strategies we can compare their results qualitatively. 

As previously demonstrated~\cite{buldyrev2010catastrophic,parshani2010interdependent}, any initial attack of an interdependent system may result in a cascading process of failures, a result of the synergy between two different effects: (a) a percolation process governed by connectivity links and (b) the failure of dependent-nodes due to a failure of their related nodes in the other network. Eventually, the cascade can either fade out, leaving a functional part of the system or alternatively, leads to complete fragmentation of the entire system. We define the size of the final largest connected component of network $A$ as $P_{\infty}$, and we study how $P_{\infty}$ changes with $p$ for different values of $q$. We also denote the size of the second-largest component of $A$ at the end of the cascading process as $P_{\infty2}$.

\FloatBarrier


\section{Analytical analysis of dependency-based attacks}

Here we present a step-by-step analytical approach for the percolation of partially interdependent networks under dependency-based targeted attacks. This can be regarded as an extension of the analytical results shown in a previous work regarding random attack~\cite{parshani2010interdependent}.

\subsection{`Dependency-first' attack}
\label{Sec:theorybeta1}

We first focus on the dependency-first attack ($\beta=1$).
After randomly attacking a fraction $1-p$ of nodes (dependency-nodes are chosen first) in $A$, there are two cases: the fraction $q$ of dependency-nodes are all removed, or not, depending on if $q>1-p$.
For either cases, we denote the fraction of remaining nodes in $A$ as $P_1^\prime=p$, and the remaining network as $A_1^\prime$.
Note that, immediately after the initial attack, the fraction of dependency nodes becomes 
\begin{eqnarray}
	q_1^\prime=
	\begin{cases}
	\frac{q-(1-p)}{p},  \quad q>1-p, \\
	0, \quad q\leq 1-p
	\end{cases}
\end{eqnarray}
Due to the initial random attack, some more nodes in $A$ will fail due to percolation. The remaining size of the giant component in network $A$ at time step $t=1$ is $P_1=P_1^\prime g_A(P_1^\prime)$ (denoted as network $A_1$). 
In fact, for the nodes in network $A$ that do not belong the giant component (totally $(1-P_1)\cdot N$ nodes), only a fraction $q_1$ of them have dependency links, where
$q_1=\frac{\mathrm{min}(q,1-p)+(p-P_1)q_1^\prime}{1-P_1}$. 
Therefore, the fraction of nodes in $B$ that fail due to inter-dependency is $q_1\cdot(1-P_1)=q_1\cdot(1-P_1^\prime g_A(P_1^\prime))$. We denote the remaining fraction of nodes in $B$ is $Q_1^\prime=1-q_1(1-pg_A(P_1^\prime))$. This remaining network is $B_1^\prime$. After that, due to percolation, the remaining giant component size in network $B$ is $Q_1=Q_1^\prime g_B(Q_1^\prime)$ (denoted as network $B_1$).

At time step $t=2$, $q_2^\prime\cdot(Q_1^\prime-Q_1)\cdot N$ more nodes in $A_1$ will fail due to interdependency, where $Q_1^\prime-Q_1$ is the fraction of percolation failures in $B$ at $t=1$, and $q_2^\prime=\frac{q-(1-Q_1^\prime)}{Q_1^\prime}$ is the fraction of dependency nodes in $B$ after attacking a fraction $1-Q_1^\prime$ of nodes on $B$. Therefore, the fraction of nodes in network $A_1$ that will be removed is $\frac{q_2^\prime}{P_1}\cdot(Q_1^\prime-Q_1)$. This is equivalent to randomly remove the same fraction of nodes from network $A_1^\prime$. Thus, the total fraction of random attack from network $A$ is $p\cdot \frac{q_2^\prime}{P_1}\cdot(Q_1^\prime-Q_1) + (1-p)=1-p(1-q_1^\prime(1-g_B(Q_1^\prime)))$. Therefore, the remaining fraction of nodes in $A$ after this attack is $P_2^\prime=p(1-q_1^\prime(1-g_B(Q_1^\prime)))$, and the giant component size in $A$ at $t=2$ is $P_2=P_2^\prime g_A(P_2^\prime)$.
Following this approach, we can obtain that at $t=2$ the remaining fraction of nodes in $B$ after the equivalent random attack is $Q_2^\prime=p(1-q_1^\prime(1-g_A(P_2^\prime)))$. After that, due to percolation, the remaining giant component size in network $B$ is $Q_2=Q_2^\prime g_B(Q_2^\prime)$.

In this way, we can obtain a general form for the giant component sizes of $A$ and $B$ at time step $t$ with $t\geq 2$:
\begin{equation}
  \begin{split}
  P_t^\prime &=p(1-q_1^\prime(1-g_B(Q_{t-1}^\prime))), \quad
  P_t=P_t^\prime g_A(P_t^\prime), \\ 
  Q_t^\prime &=p(1-q_1^\prime(1-g_A(P_t^\prime))), \quad
  Q_t=Q_t^\prime g_B(Q_t^\prime).
  \end{split}
  \label{Eq:Cascade}
\end{equation}
At the end of the cascading process, where $t\to \infty$, we have $P_{t+1}^\prime=P_t^\prime$, and $Q_{t+1}^\prime=Q_t^\prime$. We denote $x=P_{t}^\prime$, and $y=Q_{t}^\prime$ at the end of the process, then we obtain a system of equations about $x$ and $y$: $x=p(1-q_1^\prime(1-g_B(y)))$, and $y=p(1-q_1^\prime(1-g_A(x)))$.

Here, the functions $g_A(x)$ and $g_B(y)$ can be obtained through a widely-used generating function approach~\cite{buldyrev2010catastrophic,parshani2010interdependent}. We denote the generating function of degree distribution of networks $A$ as $G_{A0}(\xi)=\sum_k P_A(k)\xi^k$. We also define the generating function of the underlying branching processes as $G_{A1}(\xi)=G_{A0}^\prime(\xi)/G_{A0}^\prime(1)$. The fraction of nodes in the giant component after randomly removing $1-p$ nodes from $A$ will be $p_A(p)=1-G_{A0}(1-p(1-f_A))$, where $f_A$ can be solved from $f_A=G_{A1}(1-p(1-f_A))$~\cite{buldyrev2010catastrophic,parshani2010interdependent}. 
For an ER network $A$, we know that $G_{A1}(\xi)=G_{A0}(\xi)=\mathrm{exp}(a(\xi-1))$, where $a$ is the mean degree of network $A$. Therefore, the equations about $x$ and $y$ become: 
\begin{eqnarray}
\begin{cases}
	x=p(1-q_1^\prime f_B), \\ 
 	y=p(1-q_1^\prime f_A), \label{Eqsforxy}
\end{cases}
\end{eqnarray} 
where $f_A=\mathrm{exp}(-ax(1-f_A))$ and $f_B=\mathrm{exp}(-bx(1-f_B))$. Finally, by substitute $x$ and $y$ into Eq.~(\ref{Eqsforxy}), we have 
\begin{eqnarray}
\begin{cases}
	f_A=e^{-ap(1-f_A)(1-q_1^\prime f_B)}, \\ 
 	f_B=e^{-bp(1-q_1^\prime f_A)(1-f_B)}, \label{EqsforfAfB1}
\end{cases}
\end{eqnarray} 
By solving this system of equations, we can obtain $f_A$ and $f_B$ for given $p$ and $q$ values. 
After that, the size of the giant component in $A$ and $B$ are $P_{\infty}=xg_A(x)=p(1-f_A)(1-q_1^\prime f_B)$, and $Q_{\infty}=yg_B(y)=p(1-q_1^\prime f_A)(1-f_B)$, respectively.

To obtain the analytical results of the transition point $p_c$ and the type of the transition, we can rewrite the equations of $f_A$ and $f_B$ as 
\begin{eqnarray}
\begin{cases}
f_A=f_A(f_B)=\frac{1}{q_1^\prime}\left(1-\frac{-\mathrm{ln}(f_B)}{bp(1-f_B)}\right), \\
f_B=f_B(f_A)=\frac{1}{q_1^\prime}\left(1-\frac{-\mathrm{ln}(f_A)}{ap(1-f_A)}\right). \label{EqsforfAfB2}
\end{cases}
\end{eqnarray}
By solving Eq.~(\ref{EqsforfAfB2}) together with $\frac{d f_A(f_B)}{d f_B}\cdot \frac{d f_B(f_A)}{d f_A}=1$, we can obtain the value of $p_c$ for the first order percolation transition, where the two curves $f_A=f_A(f_B)$ and $f_B=f_B(f_A)$ become tangent at a point of $(f_A,f_B)$ with $0<f_A<1$ and $0<f_B<1$.
On the other hand, By solving Eq.~(\ref{EqsforfAfB2}) and $f_A=1$ (or $f_B=1$), we can find the transition point $p_c$ for the second order transition. 

In this way, the cascading process of failures, the final giant component size $P_{\infty}$ and $Q_{\infty}$, as well as the critical point $p_c$ can be solved numerically using the above approach.

\subsection{`Dependency-last' attack}

Here we show the analytical results for the dependency-last attack ($\beta=-1$).
As in the case of dependency-first attack, at $t=1$, a fraction $1-p$ of nodes in $A$ are initially removed. The fraction of remaining nodes is $P_1^\prime=p$, and then the giant component size in $A$ is $P_1=P_1^\prime g_A(P_1^\prime)$.

Here we use a similar approach to the case with $\beta=1$ to find the equivalent amount of random attack and the giant component size at each time step $t$. 
For dependency-last attack, the expressions of $q_1^\prime$ and $q_1$ will be different with those for $\beta=1$. Here we can find that 
\begin{eqnarray}
\begin{cases}
q_1^\prime=1, \  q_1=\frac{q-P_1}{1-P_1}, \quad q>p, \\
q_1^\prime=q/p, \  q_1=\frac{(p-P_1)q_1^\prime}{1-P_1}, \quad q\leq p. 
\end{cases}
\end{eqnarray}
For network $B$ at $t=1$, we have $Q_1^\prime=1-q_1(1-pg_A(P_1^\prime))$, and $Q_1=Q_1^\prime g_B(Q_1^\prime)$.
 
After that, at time step $t\geq 2$, we can obtain: 
\begin{equation}
P_t^\prime=p(1-q_1^\prime(1-g_B(Q_{t-1}^\prime))), \quad
P_t=P_t^\prime g_A(P_t^\prime), 
\end{equation}
\begin{eqnarray}
Q_t^\prime=
\begin{cases}
  1-q+pg_A(P_t^\prime),  \quad q>p, \\
  1-pq_1^\prime (1-g_A(P_t^\prime)),  \quad q\leq p, \\
\end{cases}
Q_t=Q_t^\prime g_B(Q_t^\prime).
\end{eqnarray}
Based on these equations, we can write the equations of $x$ and $y$ at $t=\infty$:
for $q>p$,
\begin{eqnarray}
\begin{cases}
 x=p(1-q_1^\prime (1-g_B(y))), \\
 y=1-q+pg_A(x),
\end{cases}
\end{eqnarray}
and for $q\leq p$,
\begin{eqnarray}
\begin{cases}
 x=p(1-q_1^\prime (1-g_B(y))), \\
 y=1-pq_1^\prime(1-g_A(x)).
 \end{cases}
\end{eqnarray}

These equations can be transformed into equations of $f_A=1-g_A(x)$ and $f_B=1-g_B(y)$: for $q>p$,
\begin{eqnarray}
\begin{cases}
 f_A=e^{-ap(1-f_A)(1-f_B)}, \\
 f_B=e^{-b(1-q+p(1-f_A))(1-f_B)}.
\end{cases}
\end{eqnarray}
For $q\leq p$, we can obtain similarly,
\begin{eqnarray}
\begin{cases}
 f_A=e^{-ap(1-f_A)(1-\frac{q}{p}f_B)}, \\
 f_B=e^{-b(1-qf_A))(1-f_B)}.
\end{cases}
\end{eqnarray}

As we have done for the dependency-first attack ($\beta=1$), for each of these cases here, we can express the equations as $f_A=f_A(f_B)$, and $f_B=f_B(f_A)$, and solve the percolation transition point $p_c$ numerically. Based on these results, we can also calculate the giant component size $P_{\infty}=xg_A(x)=p(1-f_A)(1-q_1^\prime f_B)$, and 
\begin{eqnarray}
Q_{\infty}=yg_B(y)=
\begin{cases}
(1-q+p(1-f_A))(1-f_B),  \quad q>p, \\
(1-pq_1^\prime f_A)(1-f_B), \quad q\leq p,
\end{cases}
\end{eqnarray}	
for given $p$ and $q$ values for the dependency-last attack.

\section{Results}

\subsection{Percolation transition due to dependency-based attacks}

In this subsection, we present analytical and simulation results of dependency-based targeted attacks on a pair of partially interdependent Erd\H{o}s-R\'{e}nyi (ER) networks. In the simulations, each network consists of $3\times10^5$ nodes with equal mean degree in both networks $k_A=k_B=4$. A random fraction $q$ of the nodes in each network are dependent upon random nodes of the other network via bidirectional dependency links (as explained above).     

\begin{figure}
 \centering
 \subfigure{
  \label{Fig:Percolationa}
  \includegraphics[width=0.32\textwidth,trim=10 10 10 10,clip=false]{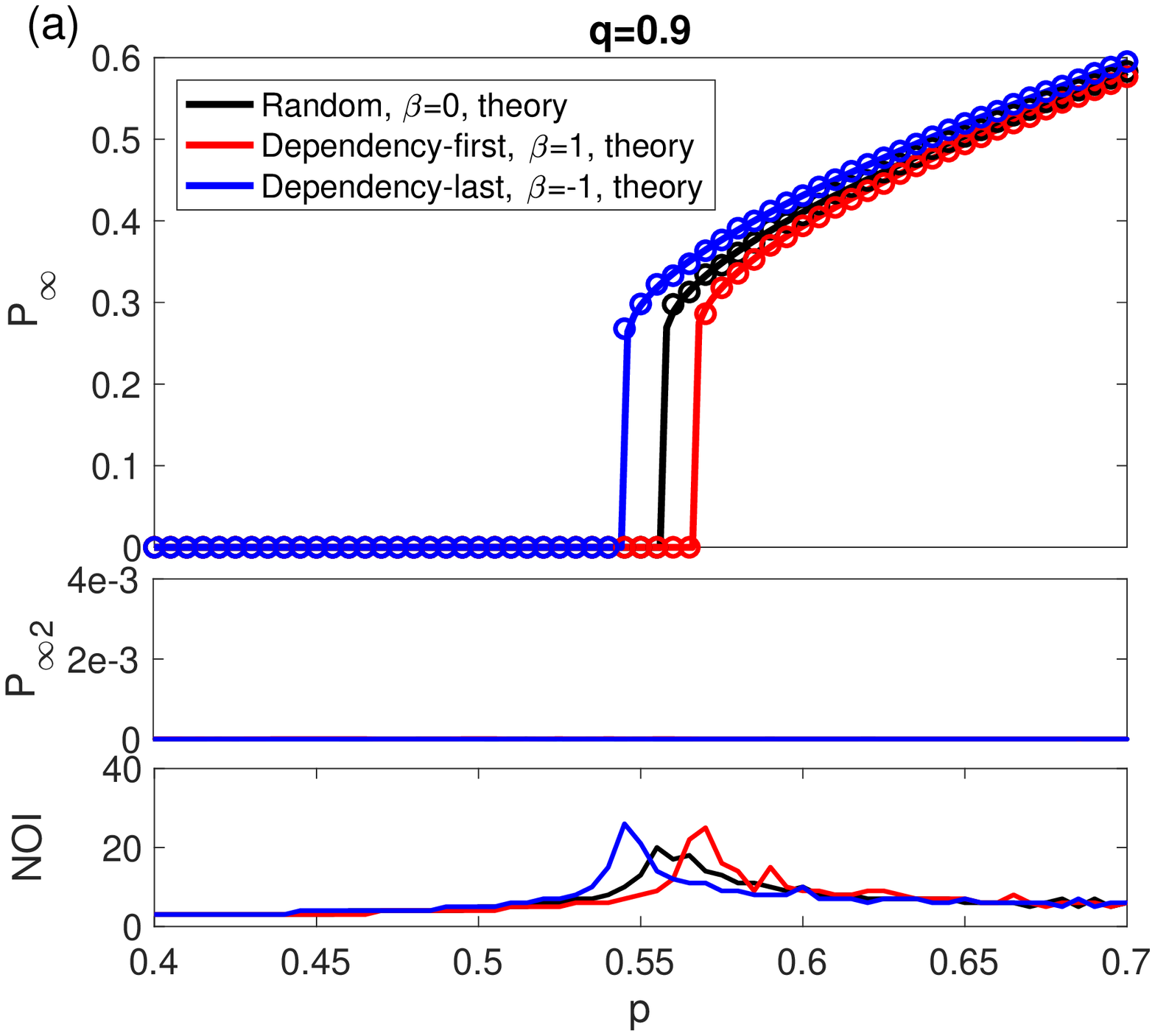}}
 \subfigure{
  \label{Fig:Percolationb}
  \includegraphics[width=0.32\textwidth,trim=10 10 10 10,clip=false]{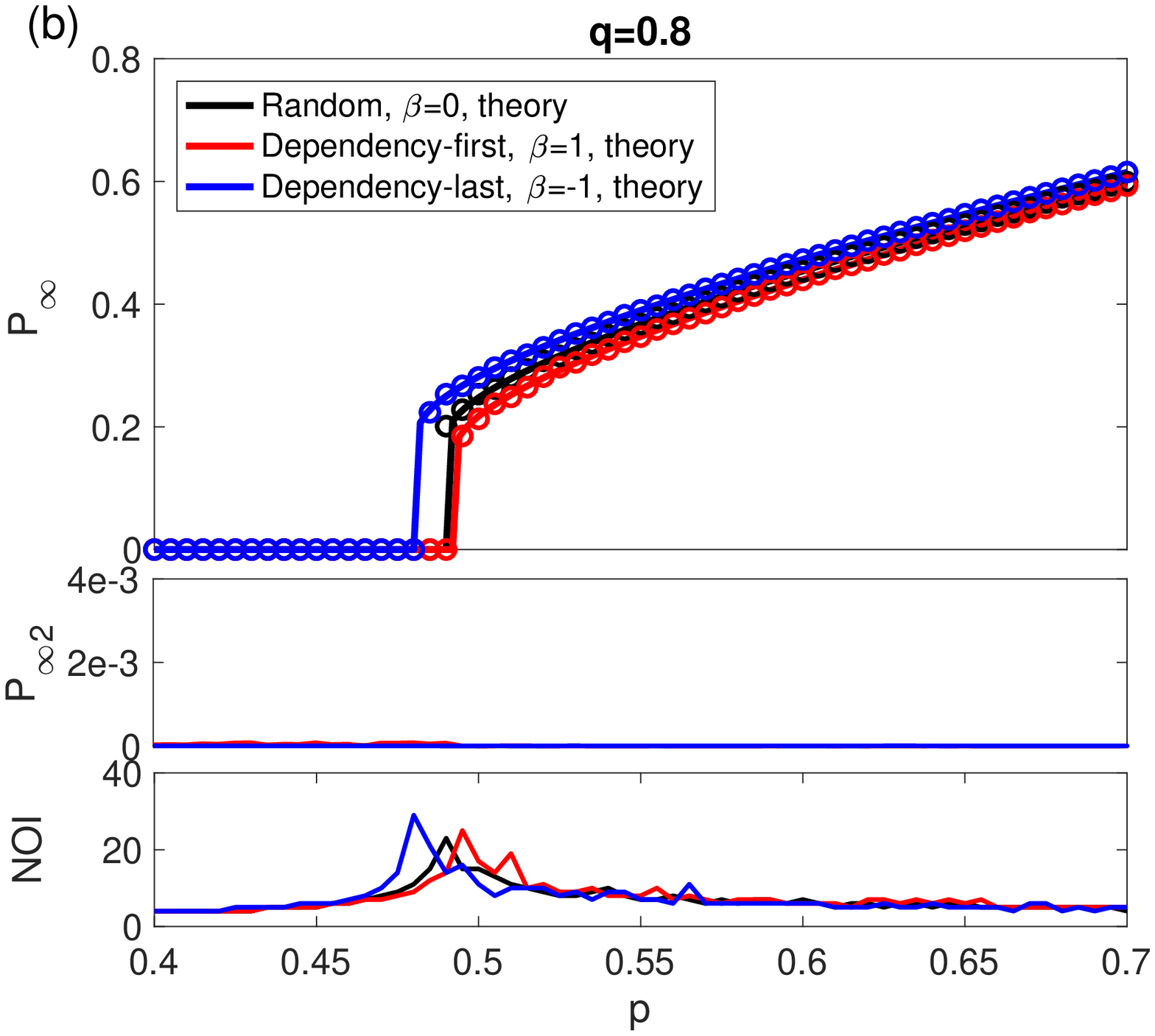}}
  \subfigure{
  \label{Fig:Percolationc}
  \includegraphics[width=0.32\textwidth,trim=10 10 10 10,clip=false]{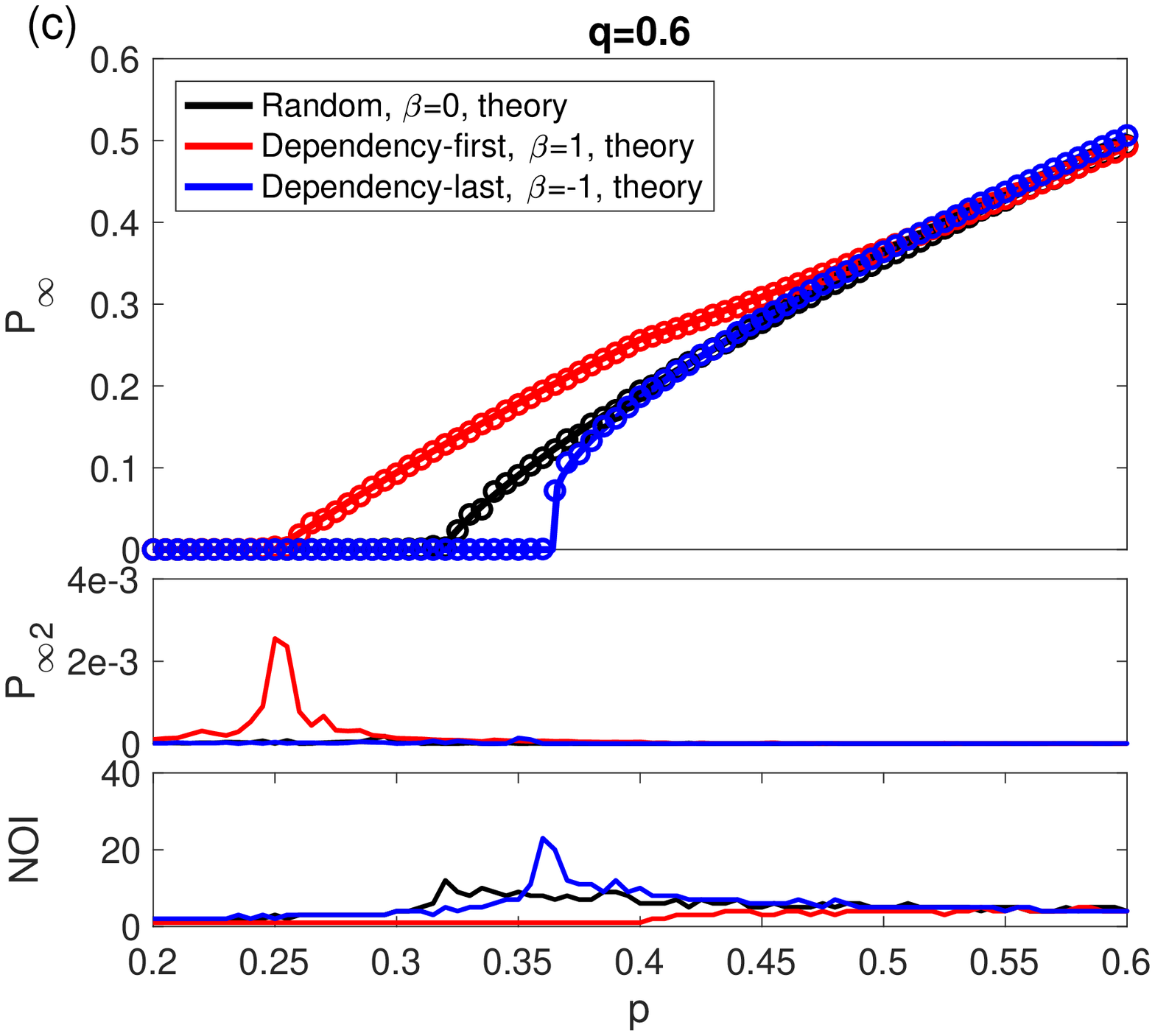}}
 \caption{Percolation transition of partially interdependent ER networks ($k_A=k_B=4$) under targeted attacks. Three levels of dependency strength are shown: (a) $q=0.9$, (b) $q=0.8$, and (c) $q=0.6$.
 Upper panels: The size of network A at steady state, $P_{\infty}$, after an initial attack of ($1-p$) of the nodes is shown versus the fraction of the remaining nodes $p$. Random attack ($\beta=0$, black) is compared with 'dependency-first' ($\beta=1$, red) and 'dependency-last' ($\beta=-1$, blue). Circles and solid lines represent the simulation and analytical results, respectively.
 Middle and lower panels show simulation results of the second-largest component of network A, $P_{\infty2}$, and the number of iterations in the cascade (NOI), respectively, versus $p$.  The numerical results represent single realization of interdependent networks, each with $N=10^6$ nodes.}
 \label{Fig:Percolation}
\end{figure}

To study the stability of interdependent networks against dependency-based attacks we analyze the percolation transition for different attack strategies. The transition is commonly represented by the percolation threshold $p_c$, and the type of the percolation transition is classified as continuous or abrupt~\cite{parshani2010interdependent}. The analytical derivation of $p_c$ is given by Eq. (\ref{EqsforfAfB2}), as shown above in Sec.~\ref{Sec:theorybeta1}. The numerical analysis is demonstrated in Fig.~\ref{Fig:Percolation} for the extreme cases of 'Dependency-first' ($\beta=1$), 'Dependency-last' ($\beta=-1$) and a random attack ($\beta=0$). At the transition point $p_c$, the size of the largest connected component of network A, $P_{\infty}$, changes from zero to non-zero (Fig.~\ref{Fig:Percolation}, top panels). In the case of a continues transition, i.e., the size of the largest-component at criticality is zero, the size of the second-largest component peaks as $p$ approaches $p_c$, as in regular percolation in a single network (Fig.~\ref{Fig:Percolation}, middle panels)~\cite{cohen2010complex,stauffer2003}. The case of abrupt transition, i.e., the size of the network at criticality is larger than zero, is characterized by a sharp peak in the number-of-iterations (NOI) in the cascade (Fig.~\ref{Fig:Percolation}, bottom panels)~\cite{hu2013percolation,Parshani2011}. The simulations results are in agreement with the analytical predictions.

As demonstrated in Fig.~\ref{Fig:Percolation}, the response of an interdependent system to an dependency-based targeted attack is dramatically affected by the coupling strength $q$. In the case of strong dependency, $q=0.9$, the system is more susceptible to 'dependency-first' attack strategy compared with 'dependency-last' attack, yielding higher percolation threshold $p_c$ (Fig.~\ref{Fig:Percolationa}). However, for weaker dependency strength $q=0.8$, $p_c$ is roughly the same in both attack strategies (Fig.~\ref{Fig:Percolationb}), while in weak coupling $q=0.6$ the system becomes more susceptible to 'dependency-last' compared with 'dependency-first' (Fig.~\ref{Fig:Percolationc}). In all cases, the random strategy represents the intermediate case between the two extreme cases. 

\begin{figure}
 \centering
 \subfigure{
  \label{Fig:Pc_vs_qa}
  \includegraphics[width=0.42\textwidth,trim=10 10 10 10,clip=false]{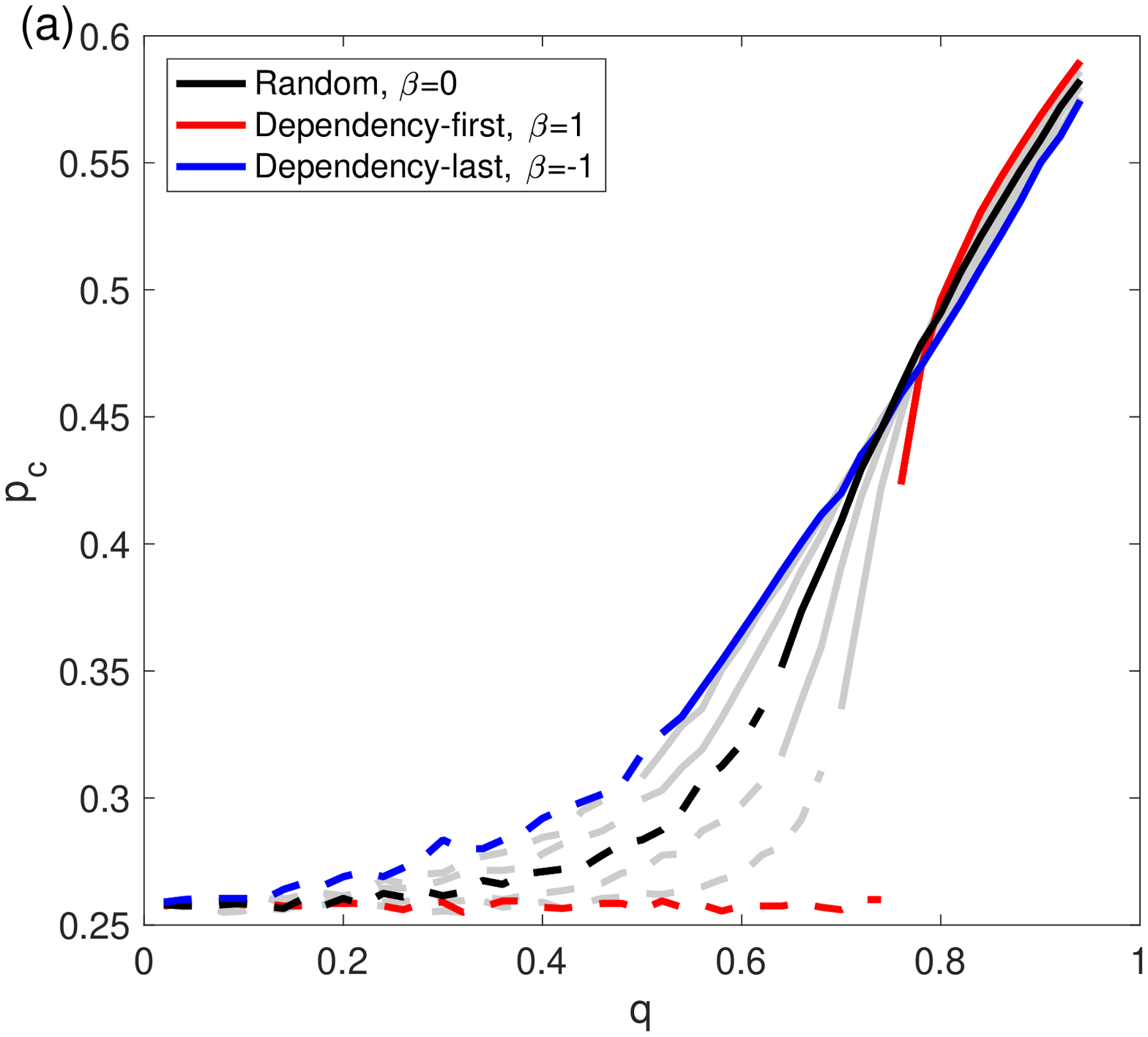}}
 \subfigure{
  \label{Fig:Pc_vs_qb}
  \includegraphics[width=0.42\textwidth,trim=10 10 10 10,clip=false]{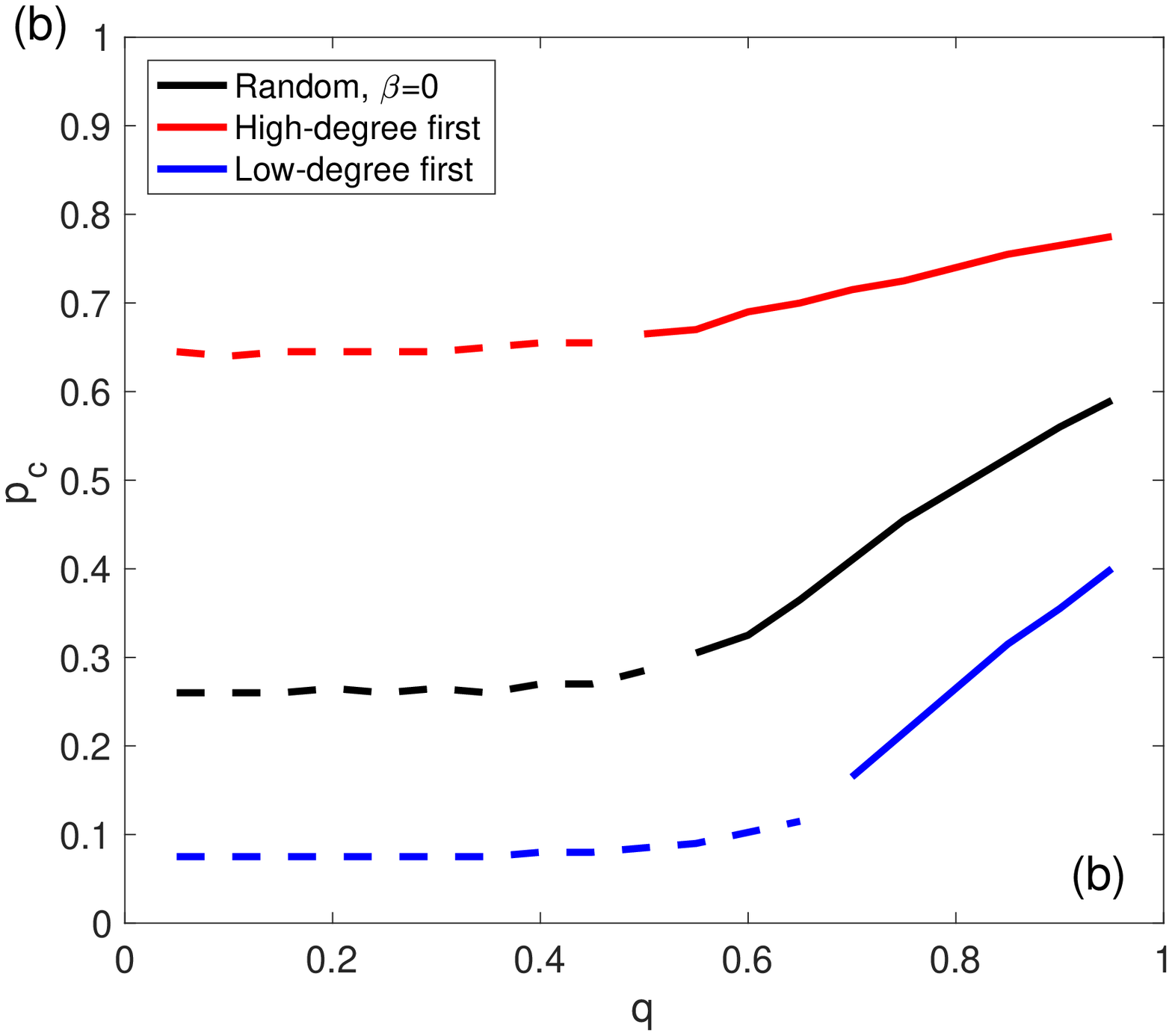}}
 \subfigure{
  \label{Fig:Pc_vs_qc}
  \includegraphics[width=0.42\textwidth,trim=10 10 10 10,clip=false]{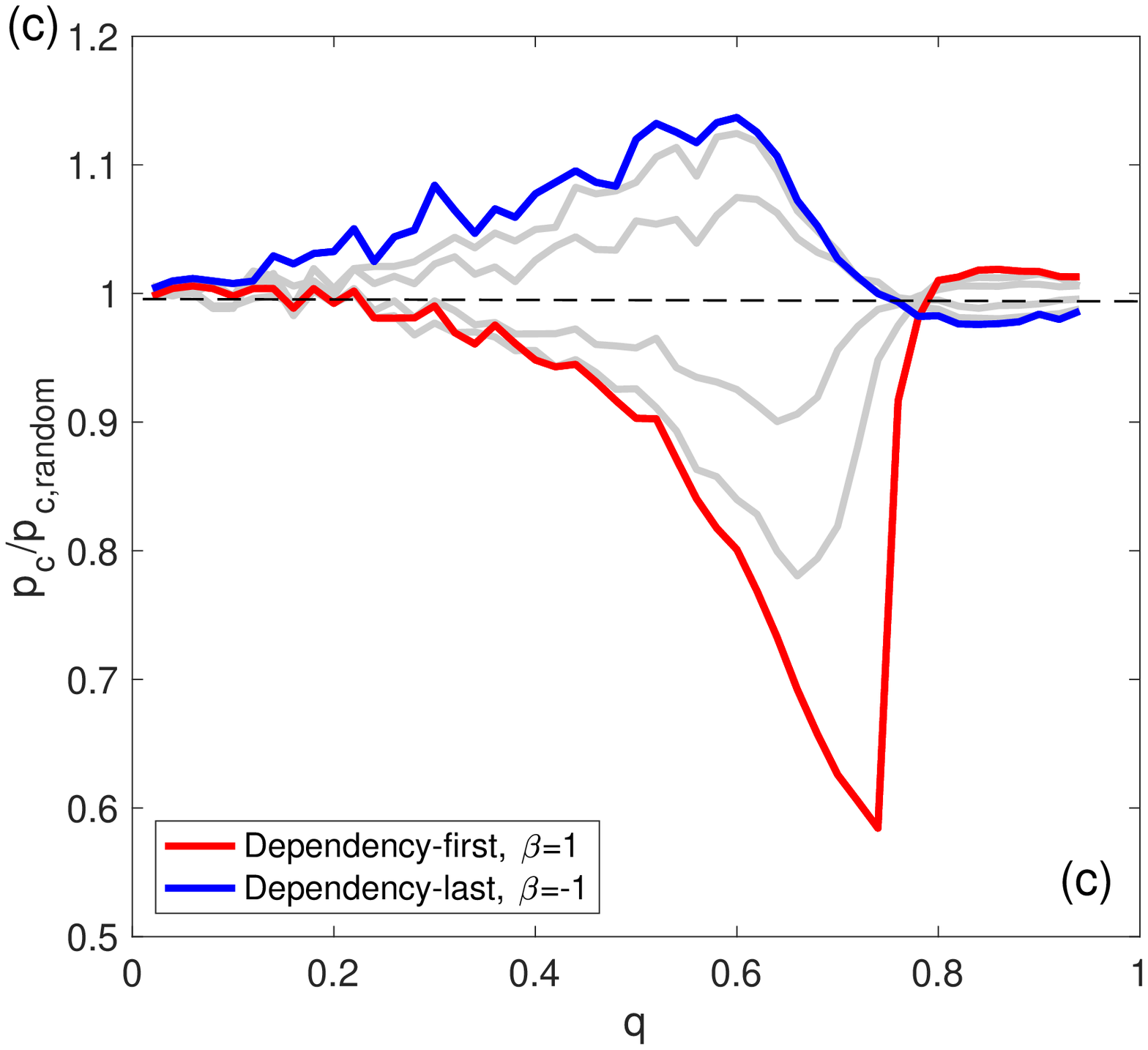}}
 \subfigure{
  \label{Fig:Pc_vs_qd}
  \includegraphics[width=0.42\textwidth,trim=10 10 10 10,clip=false]{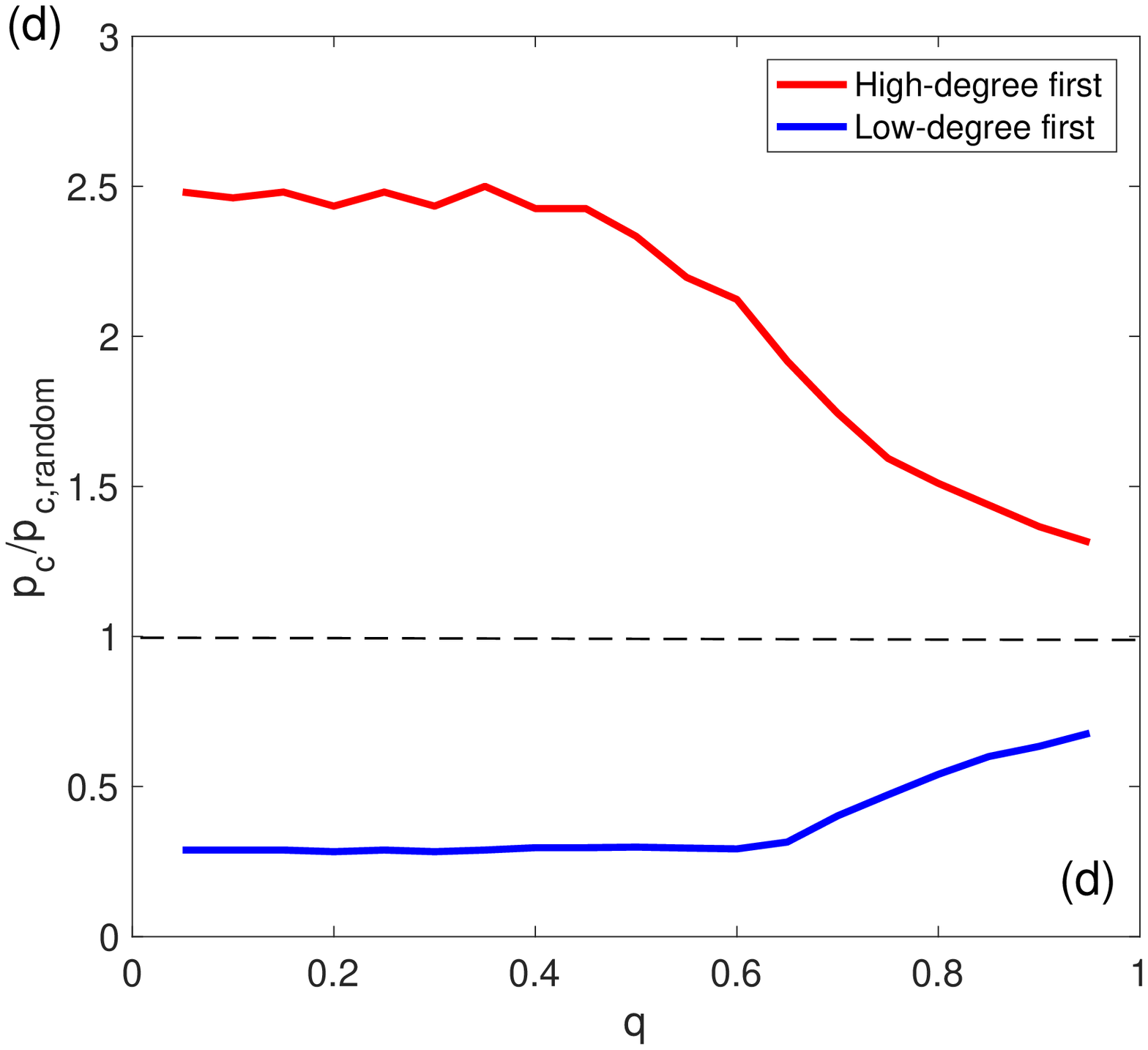}}
\caption{The effect of coupling strength $q$ on the stability of interdependent networks under dependency-based and degree-based attacks. (a) Numerical results of $p_c$ versus $q$, for partially interdependent ER networks ($k_A=k_B=4$), each with $N=3\times 10^5$ nodes. Results represent an average over 10 realizations. Random attack (black) is compared with dependency-first ($\beta=1$, red) and dependency-last ($\beta=-1$, blue) targeted attacks. 
Solid lines and dashed lines indicate first order and second order transitions, respectively, for each curve. Grey lines represent four intermediate cases $\beta=\pm 1/3$, $\beta=\pm 2/3$ (grey lines).
The red dashed line corresponds to the region where all dependency nodes are initially attacked. 
(b) The same as (a) for the cases of high-degree first (red) and low-degree first (blue) targeted attacks. 
The ratio between the critical threshold $p_c$ of each of the attacks and the one for same size random attack, $p_c/p_{c,random}$, (c) for dependency-based attacks and (d) for degree-based attacks.}
 \label{Fig:Pc_vs_q}
\end{figure}

To study the role of the dependency strength in determining the stability of interdependent networks against dependency-based attacks, we systematically analyze the percolation threshold for different attack strategies for various values of coupling strength $q$.
In Fig.~\ref{Fig:Pc_vs_qa}, we show $p_c$ as a function of $q$ for dependency-based attacks with $\beta=\pm 1, \pm2/3, \pm1/3$, as well as for random attacks ($\beta=0$).
There is a general trend common for all attack strategies. Increasing the coupling strength leads to higher $p_c$, and the transition type change from continuous transition to abrupt transition, as previously shown for random attack~\cite{parshani2010interdependent}. In respect to this trend, Fig.~\ref{Fig:Pc_vs_qc} compares the transition due to each of the different dependency-based attack strategies with the transition due to random attack over the same number of nodes. We find that for each of the 'dependency-biased' attack strategies ($\beta>0$), there is coupling strength $q^{*(\beta)}$, such that for $q>q^{*(\beta)}$ the critical threshold $p_c$ is higher than random, while for $q<q^{*(\beta)}$, the critical threshold $p_c$ is lower. In contrast, for the 'non-dependency-biased' attack strategies ($\beta<0$), for $q>q^{*(\beta)}$ the critical threshold $p_c$ is lower than random, while for $q<q^{*(\beta)}$, the critical threshold $p_c$ becomes higher than for the case of random attack. As explained above, this is the result of the effective change in the dependency strength after 'dependency-based' attacks.

In contrast, the effect of 'degree-based' attacks that target nodes based on the number of neighbors within each network is qualitatively the same for all dependency strengths. As shown in Figs.~\ref{Fig:Pc_vs_qb} and ~\ref{Fig:Pc_vs_qd}, $p_c$ of 'high-degree-first' is higher compared with random and $p_c$ of 'high-degree-first' is lower compared with random, for any value of dependency strength $q$.

\FloatBarrier

\subsection{The special cascading dynamics due to dependency-based attacks}

\begin{figure}
 \centering
 \subfigure{
  \label{fig5a}
  \includegraphics[width=0.42\textwidth,trim=10 10 10 10,clip=false]{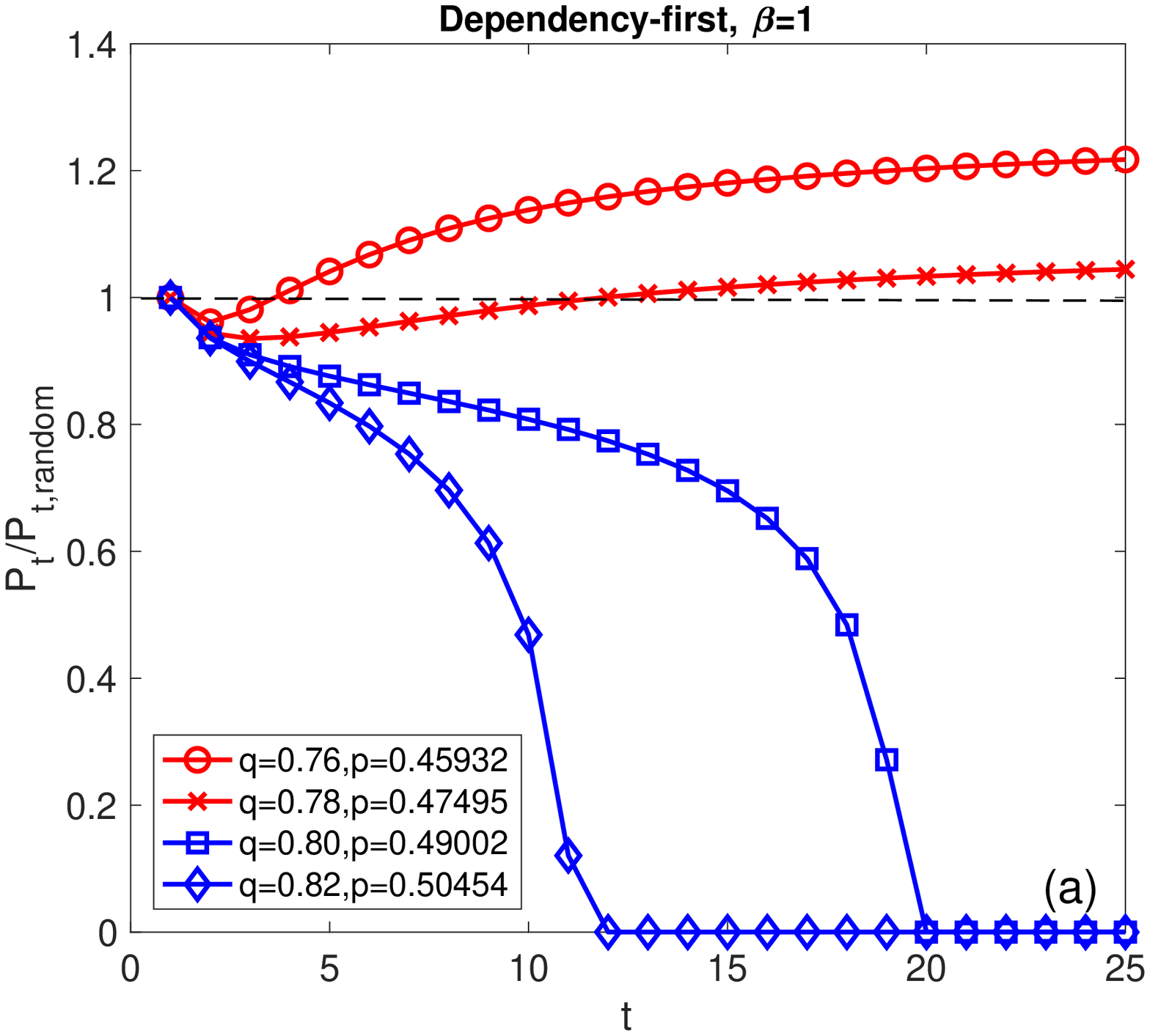}}
 \subfigure{
  \label{fig5b}
  \includegraphics[width=0.42\textwidth,trim=10 10 10 10,clip=false]{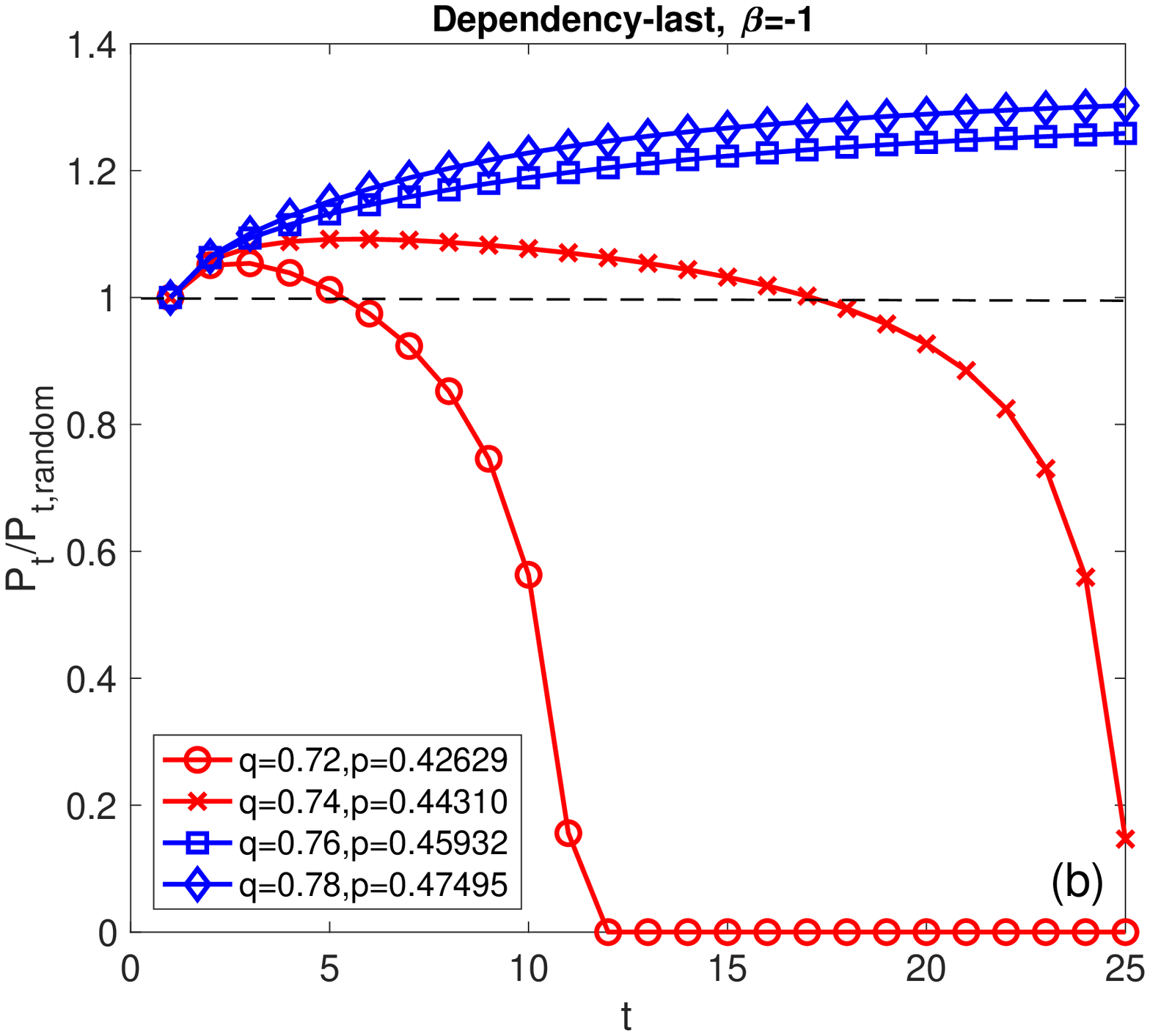}}
 \caption{(a) The ratio between the giant component size of $A$ under dependency-first attack ($\beta=1$) and that under random attack, $P_{t}/P_{t,random}$ VS $t$, analytical results. Partially interdependent ER networks. Here $k_A=k_B=4$, $q$ changes from $0.76$ to $0.82$, and $p$ is taken as the critical value $p_c$ in theory.
(b) The same as (a), but for the dependency-last attack ($\beta=-1$).
}
 \label{fig5}
\end{figure}

In order to understand the counter-intuitive effect of the dependency strength in dependency-based targeted attacks (shown in Fig.\ref{Fig:Pc_vs_q}), we explore the dynamics of the cascading process near $q^{*(\beta)}$. 
Figure~\ref{fig5a} shows, for each step of the cascade, the ratio between the sizes of network A under dependency-first attack and under same-size-random-attack, $P_{t}/P_{t,random}$, calculated from Eq.(\ref{Eq:Cascade}). 
As shown in Fig.~\ref{Fig:Pc_vs_qc}, for the case of $\beta=1$ the crossover is at $q^{*(\beta=1)}\approx 0.79$. Accordingly, in Fig.~\ref{fig5a} we show the cascading process of models with values of $q$ ranging from 0.76 to 0.82, where for each model the initial attack size, ($1-p^{\beta=1}$), is chosen as the critical value for the random attack, i.e., $p^{\beta=1}=p_c^{\beta=0}$, as detailed in the figure legend. In the case of $q>q^{*(\beta=1)}$, i.e., $q=0.80$ and $q=0.82$, the sequence of $P_{t}/P_{t,random}$ monotonically decreases. This behavior is rather intuitive since a dependency-first attack causes an increased immediate damage compared with a random attack, which in turn leads to a series of larger failures and eventually a complete collapse of the system. However, for $q<q^{*(\beta)}$, i.e., $q=0.76$ and $q=0.78$, the sequence of $P_{t}/P_{t,random}$ is non-monotonic. The immediate damage right after the initial attack is indeed larger than random, that is, $P_{t}/P_{t,random} < 1$, but the following cascading failures become smaller than random until eventually the system reaches a steady state with a network size which is larger than same-size-random-attack, $P_{\infty}/P_{\infty,random} > 1$.

Similarly, in the case of dependency-last attacks ($\beta=-1$) the dynamics of the cascading failures dramatically change for different coupling strengths, as shown in Fig.~\ref{fig5b}.
In contrast with dependency-first attacks, the initial impact of dependency-last attacks is smaller than random failures, yielding larger network size ($P_{\infty}/P_{\infty,random} > 1$) right after the initial attack. This is expected, as the attack is focus mainly on non-dependent nodes. However, in the cases where $q<q^{*(\beta)}$ ($q=0.72$ and $q=0.74$) the damage in each step of the cascade gradually increases until the system completely collapses. 

These counter-intuitive behaviors stem from the role the dependency links play throughout the process. In the case of 'dependency-first' attack, the immediate impact is indeed larger than random failures, However, in the long term, removing the dependent nodes also decreases the coupling between the networks leading to suppressed propagation of the damages. As an extreme example, attacking all the dependent nodes at the initial step, i.e., 'dependency-first' attack with $1-p=q$, will cause large immediate damage but no additional cascading failures. In contrast, 'dependency-last' attacks minimize the immediate failures but effectively increases the coupling between the networks in the following steps. These results suggest that the effective coupling strength during the process of the cascading failures may have a crucial role in determining the overall resilience of the system.
\FloatBarrier

\section{Conclusion}

In summarize, we analytically and numerically investigated the percolation transitions in partially interdependent ER networks under targeted attack based on dependency links. We show that a counter-intuitive vulnerability arises due to the role the dependency links play throughout the cascading process, where for a wide range of dependency strength, an attack strategy that avoids the dependent nodes may lead to more damage compared with an attack that focuses on them. A general conclusion from these results is that the effect of dependency-based targeted attacks should not be considered solely by their immediate impact but the evolving process of cascading failures should be taken into account. 

\section{Acknowledgements}
A.B. thanks Azrieli Foundation for supporting this research.

\newpage
\bibliography{References}

\end{document}